\documentclass[twocolumn,journal,10pt,letterpaper,final]{IEEEtran}
\usepackage[noadjust]{cite}
\usepackage{graphicx,color}
\usepackage[dvipsnames]{xcolor}
\usepackage{amsmath,amsbsy,amsfonts,amssymb}
\usepackage{mathrsfs}
\usepackage{mathtools}
\mathtoolsset{showonlyrefs}
\ifCLASSOPTIONcompsoc
\usepackage[caption=false,font=normalsize,labelfont=sf,textfont=sf]{subfig}
\else
\usepackage[caption=false,font=footnotesize]{subfig}
\fi

\usepackage{acro}
\DeclareAcronym{AWGN}{short = AWGN ,long = additive white gaussian noise}
\DeclareAcronym{AoI}{short = AoI ,long = age of information}
\DeclareAcronym{CDF}{short = CDF ,long = cumulative distribution function}
\DeclareAcronym{CRA}{short = CRA ,long = contention resolution ALOHA}
\DeclareAcronym{CRDSA}{short = CRDSA ,long = contention resolution diversity slotted ALOHA}
\DeclareAcronym{CSA}{short = CSA ,long = coded slotted ALOHA}
\DeclareAcronym{C-RAN}{short = C-RAN ,long = cloud radio access network}
\DeclareAcronym{DAMA}{short = DAMA ,long = demand assigned multiple access}
\DeclareAcronym{DSA}{short = DSA ,long = diversity slotted ALOHA}
\DeclareAcronym{eMBB}{short = eMBB ,long = enhanced mobile broadband}
\DeclareAcronym{FEC}{short = FEC ,long = forward error correction}
\DeclareAcronym{GEO}{short = GEO ,long = geostationary orbit}
\DeclareAcronym{IC}{short = IC ,long = interference cancellation}
\DeclareAcronym{IoT}{short = IoT ,long = internet of things}
\DeclareAcronym{IRSA}{short = IRSA ,long = irregular repetition slotted ALOHA}
\DeclareAcronym{LEO}{short = LEO ,long = low Earth orbit}
\DeclareAcronym{M2M}{short = M2M ,long = machine-to-machine}
\DeclareAcronym{MAC}{short = MAC ,long = medium access}
\DeclareAcronym{MPR}{short = MPR ,long = multi-packet reception}
\DeclareAcronym{MTC}{short = MTC ,long = machine-type communications}
\DeclareAcronym{mMTC}{short = mMTC ,long = massive machine-type communications}
\DeclareAcronym{PDF}{short = PDF ,long = probability density function}
\DeclareAcronym{PER}{short = PER ,long = packet error rate}
\DeclareAcronym{PLR}{short = PLR ,long = packet loss rate}
\DeclareAcronym{PMF}{short = PMF ,long = probability mass function}
\DeclareAcronym{RA}{short = RA ,long = random access}
\DeclareAcronym{SA}{short = SA , long = slotted ALOHA}
\DeclareAcronym{SIC}{short = SIC ,long = successive interference cancellation}
\DeclareAcronym{SIR}{short = SIR ,long = signal to interference ratio}
\DeclareAcronym{SNIR}{short = SNIR ,long = signal-to-noise and interference ratio}
\DeclareAcronym{SINR}{short = SINR ,long = signal-to-interference and noise ratio}
\DeclareAcronym{SNR}{short = SNR ,long = signal-to-noise ratio}
\DeclareAcronym{TDM}{short = TDM ,long = time division multiplexing}

\begin{document}

\title{\huge Average Age of Information of\\ Irregular Repetition Slotted ALOHA}
\author{Andrea Munari, Alexey Frolov
\thanks{
A. Munari is with Institute of Communications and Navigation of the German Aerospace Center (DLR), 82234 Wessling, Germany (e-mail: \mbox{Andrea.Munari@dlr.de}).}
\thanks{A. Frolov is with the Center for Computational and
Data-Intensive Science and Engineering, Skolkovo Institute of Science and
Technology, 121205 Moscow, Russia (e-mail: al.frolov@skoltech.ru).} 
\thanks{The research of A. Frolov was carried at Skoltech and supported by the Russian Science Foundation (project no. 18-19-00673).}
}
\maketitle
\thispagestyle{empty} \setcounter{page}{0}
\vspace{-3em}

\begin{abstract}
Flanking traditional metrics such as throughput and reliability, \ac{AoI} is emerging as a fundamental tool to capture the performance of IoT systems. In this context, we focus on a setup in which a large number of nodes attempt delivery of time-stamped updates to a common destination over a shared channel, and investigate the ability of different grant-free access strategies to maintain fresh information at the receiver. Specifically, we derive for the first time an exact closed-form expression of the average \ac{AoI} achieved by \ac{IRSA}, and compare its performance to that of a slotted ALOHA approach. Our analysis reveals the potential of modern random access schemes, and pinpoints some fundamental trade-offs, providing useful hints for proper system design.
\end{abstract}

\pagestyle{empty}

\newtheorem{prop}{Proposition}
\newtheorem{lemma}{Lemma}

\newcommand{\pr}{\ensuremath{\mathbb P}}
\newcommand{\expOp}{\ensuremath{\mathbb E}}

\newcommand{\tru}{\ensuremath{\mathsf S}}
\newcommand{\truSA}{\ensuremath{\mathsf S_{\mathsf{sa}}}}
\newcommand{\psSA}{\ensuremath{\mathsf p_{s,\mathsf{sa}}}}
\newcommand{\load}{\ensuremath{\mathsf G}}

\newcommand{\nodes}{\ensuremath{\mathsf n}}
\newcommand{\slots}{\ensuremath{\mathsf m}}
\newcommand{\pAct}{\ensuremath{\pi_{\mathsf a}}}
\newcommand{\pUpdate}{\ensuremath{\nu}}
\newcommand{\psucc}{\ensuremath{\mathsf p_{\mathsf s}}}
\newcommand{\ploss}{\ensuremath{\mathsf p_{ l}}}
\newcommand{\plosswf}{\ensuremath{\mathsf p_{l,wf}}}
\newcommand{\plossef}{\ensuremath{\mathsf p_{l,ef}}}

\newcommand{\tStamp}{\ensuremath{Y^{(i)}}}
\newcommand{\tGen}{\ensuremath{X^{(i)}}}

\newcommand{\age}{\ensuremath{\delta}}
\newcommand{\agei}{\ensuremath{\age^{(i)}}}
\newcommand{\Agei}{\ensuremath{\Delta^{(i)}}}
\newcommand{\AoI}{\ensuremath{\Delta}}
\newcommand{\AoIIRSA}{\ensuremath{\AoI_{\mathsf{irsa}}}}
\newcommand{\AoISA}{\ensuremath{\AoI_{\mathsf{sa}}}}
\newcommand{\pv}{\ensuremath{\mathsf P_v}}

\newcommand{\actNodesFrame}{\ensuremath{N_a}}
\newcommand{\replicas}{\ensuremath{\ell}}
\newcommand{\tSlot}{\ensuremath{T_{\mathsf s}}}
\newcommand{\tFrame}{\ensuremath{T_{\mathsf f}}}
\section{Introduction} 
\label{sec:intro}

\IEEEPARstart{I}{nformation} freshness is gaining growing attention as a valuable tool to gauge the performance of wireless communications systems. Complementing traditional metrics such as spectral efficiency and latency, such a notion captures the importance of keeping information available at a receiver on a quantity of interest as up-to-date as possible. This aspect is paramount, among others, in environmental monitoring, vehicle and asset tracking, control of dynamic systems, and, more generally, in all \ac{IoT} applications which aim at offering a fresh view on a monitored process.

The issue was formally tackled at first in the context vehicular communications, with the introduction of the \emph{\acf{AoI}} metric  \cite{Kaul11_SECON,Kaul11_Globecom}, tracking the lag between current time and time-stamp of the last received status update from a source. Since then, a number of notable results have been derived, and a solid understanding of some fundamental trade-offs regulating \ac{AoI} in point-to-point links has been achieved, see e.g. \cite{Yates19_TIT,Durisi19_JSAC}, as well as \cite{Modiano19_book} and references therein.

In parallel, research has started focusing on the characterisation of information age in wireless networks. Along this line, preliminary insights have been derived, e.g. \cite{Yates17:AoI_SA,Modiano18_AoI,Ephremides19_Infocom,Ephremides20_CSMA}, highlighting how medium access policies play a key role in determining information freshness. Indeed, competing interests emerge when multiple status reporting devices share a common channel, as maintaining an up-to-date knowledge on the state of a terminal reduces resources available to others. This becomes especially crucial for a wide class of \ac{IoT} applications, which foresee a massive number of terminals transmit, over the same bandwidth, short updates in a sporadic and possibly unpredictable fashion, e.g. driven by measurements of physical quantities. In this context, the random activation pattern of  nodes renders scheduled transmission policies impractical, and uncoordinated grant-free strategies are the reference choice. Notably, however, although a large fraction of  commercial \ac{mMTC} solutions such as LoRaWAN and Sigfox do rely on the simplest ALOHA protocol, their design is still mainly carried out aiming at classical performance metrics. From this standpoint, gathering a clear understanding of how \ac{AoI} behaves in such system can pave the road to new design paradigms and to the identification of \ac{IoT}-targeted optimisation criteria. 
Important steps along this direction have been taken in \cite{Yates17:AoI_SA,Modiano18_AoI,Yates20_arXiv}, where closed form expressions for the average \ac{AoI} of slotted and unslotted ALOHA were derived and compared to the performance of scheduled multiple access. Similarly, variations of the scheme that allow nodes to selectively transmit packets more likely to be valuable in terms of information freshness have been proposed in \cite{Shirin19_arXiv,Soung20_arXiv}.

In turn, the growth of \ac{IoT} applications has recently led to the design of a novel family of protocols, referred to as \emph{modern random access} \cite{Berioli2016}. Such solutions foresee nodes transmit multiple copies of their packets over a frame of predefined duration according to a properly tuned probability distribution, and adopt advanced signal processing techniques at the receiver to recover information \cite{Paolini15:TIT_CSA}. Constructively embracing interference, this approach attains a throughput comparable to that of coordinated schemes even in a fully grant-free setup, rendering modern random access very appealing for \ac{mMTC} both in the terrestrial and in the satellite domain (where some protocols have already found their way into commercial standards). Despite their potential in terms of spectral efficiency, however, the behaviour of such schemes with respect of information freshness has not yet been investigated. 

In this paper we start to bridge such a gap, providing an exact characterisation of the average \ac{AoI} achievable when nodes access the channel following \acl{IRSA} (IRSA) \cite{Liva11:IRSA}. Compact and insightful closed form expressions are derived, enabling a direct comparison with the performance of a plain slotted ALOHA policy. Our results reveal how \ac{IRSA}  can significantly outperform traditional random access strategies in a wide range of operating conditions, further buttressing its use in \ac{mMTC} applications. Moreover, non-trivial trade-offs emerge, as the length of the frame over which \ac{IRSA} is operated is shown to play a key role in determining performance. The presented analysis offers in this sense a useful tool for proper system design.

\section{System Model and Preliminaries}
\label{sec:sysModel}

Let us focus on a system where \nodes\ users share a wireless channel to communicate with a common destination (or sink). Time is divided in slots of unitary duration, set to fit the transmission of a single packet. All terminals are assumed to be synchronised to this pattern and, at any slot, each of them independently becomes active with probability \pAct, producing a time-stamped status update for the sink. Such a model is for instance representative of sensor nodes, which track the evolution of a physical quantity and notify the sink of updates by sending a packet containing a field that specifies the time at which a new measurement was taken. 

Implementing a grant-free approach, active devices access the channel in an uncoordinated fashion, and attempt delivery of information following one of two strategies that will be described later in this section. In turn, the destination keeps track of incoming reports, and stores the latest successfully received update together with its generation time for all the terminals. 

In this setting, we are interested in gauging the ability of the sink to maintain up-to-date information for every node. To this aim, let us denote by $\tStamp(t)$ the timestamp of the last received update from terminal $i$ as of time $t$, and define the current age of information for the device as
\begin{equation}
    \agei(t) := t - \tStamp(t).
\end{equation}
As exemplified by Fig.~\ref{fig:timeline_AoI}, $\agei(t)$ grows linearly over time in the absence of updates from terminal $i$. Conversely, as soon as a packet sent by the node is decoded, the sink accordingly refreshes $i$'s state, and $\agei(t)$ is reset to the difference between the current time and the time at which the incoming update was generated. Following this notation, the average age of information for node $i$ can be defined as
\begin{equation}
    \Agei := \lim_{\tau \rightarrow \infty} \frac{1}{\tau} \int_{0}^{\tau} \agei(t)\, dt.
\label{eq:def_av_aoi_node}
\end{equation}
We shall assume that this limit exists, and will evaluate system performance by computing the \emph{network age of information} $\AoI$, formally introduced as
\begin{equation}
    \AoI := \frac{1}{\nodes} \sum_{i=1}^{\nodes} \Agei.
\end{equation}

In order to understand the effectiveness of modern random access schemes in granting low \ac{AoI} for \ac{mMTC} applications, we consider two distinct medium sharing policies: \ac{SA} and \ac{IRSA}. In both cases, a collision channel model is assumed, so that a packet received without interference is always decoded, whereas no information can be retrieved from a slot as long as it sees the superposition of two or more transmissions. Moreover, we characterise the efficiency of the protocols considering the \emph{aggregate throughput} \tru, defined as the average number of packets per slot decoded at the receiver.

\emph{\Acf{SA}:} Offering a relevant benchmark in view of its widespread application, the classical \ac{SA} paradigm \cite{Abramson77:PacketBroadcasting} represents the simplest variation of grant-free protocols. When implementing this approach, a node that becomes active at the beginning of a slot immediately transmits the generated update to the sink, regardless of the activity of its peers. The number of terminals accessing the medium over a slot of interest follows then a binomial distribution of parameters $(\nodes,\pAct)$, and a device is decoded at the receiver with probability \mbox{$(1- \pAct)^{\nodes-1}$}, i.e. if no other packet was concurrently sent. Neither feedback from the destination nor retransmission policies are considered, so that delivery of each update is attempted only once. Accordingly, the aggregate throughput \tru\ readily evaluates to
\begin{equation}
    \tru = \nodes \,\pAct (1- \pAct)^{\nodes-1}.
    \label{eq:truSA}
\end{equation}
Finally, we observe that the successful reception of a node's packet always resets its current \ac{AoI} to $1$, as no time other than the transmission latency elapses between the generation of the update and its reception.

\begin{figure*}[!th]
    \centering
    \subfloat[nodes' activation pattern, frame $k$]{
    \includegraphics[width=.83\columnwidth]{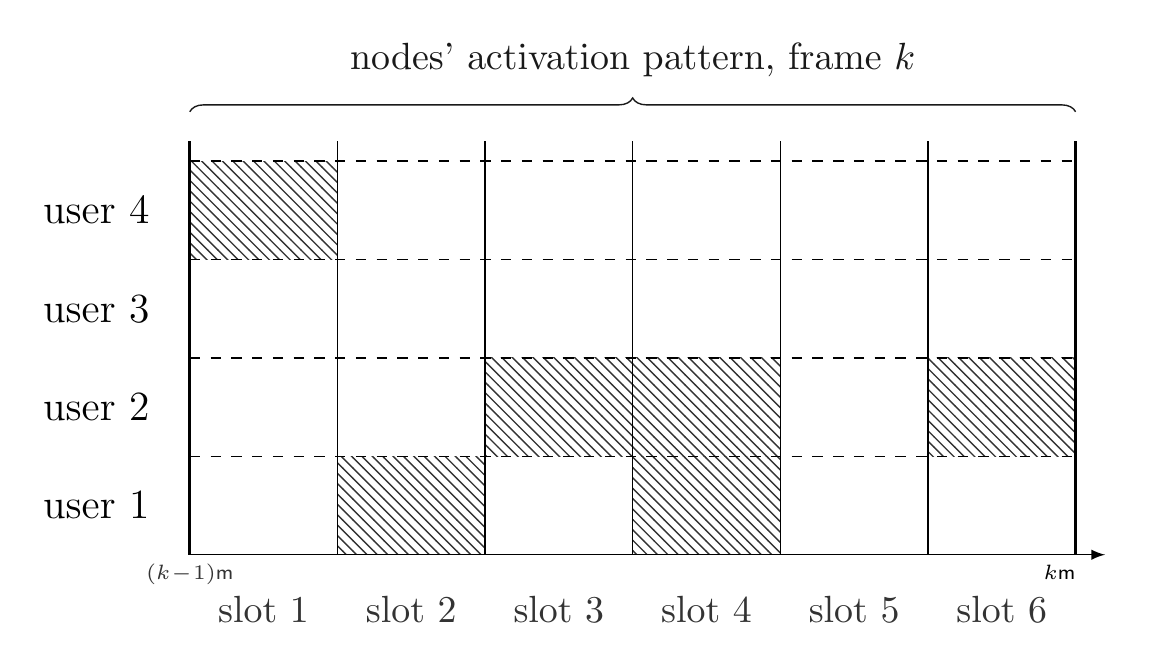}
    \label{fig:irsa_timeline_tx}
    }\hspace{2em}
    \subfloat[nodes' transmission pattern, frame $k+1$]{
    \includegraphics[width=.84\columnwidth]{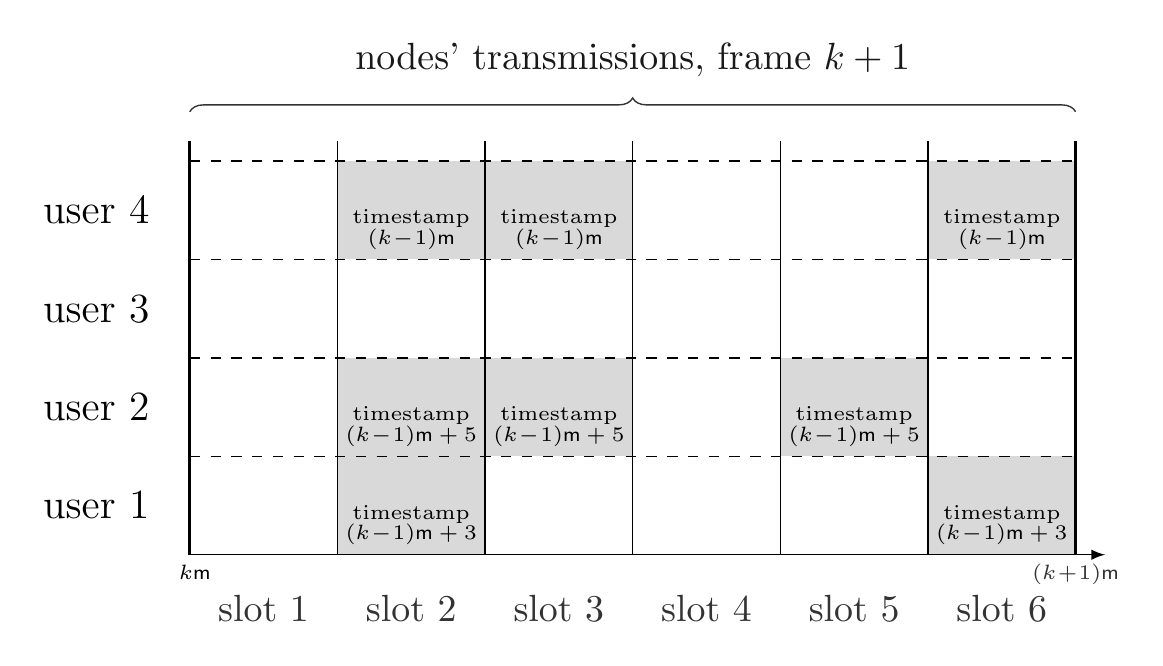}
    \label{fig:irsa_timeline_rx}
    \vspace{-.3em}
    }
    \caption{Example of operation for the IRSA protocol. On the left handside, the users' activation pattern during frame $k$ is reported (filled rectangles indicate slots in which a node generated an update). On the right handside, nodes that became active at least once during frame $k$ transmit a random number of replicas of their update over frame $k+1$. Each copy contains the same timestamp, relative to the last generated update over frame $k$.}
    \label{fig:irsa_timeline}
\end{figure*}

\emph{\Acf{IRSA} \cite{Liva11:IRSA}:} In this case, channel access is organised over frames composed by \slots\ slots each. All users are frame-synchronised, and the first transmission opportunity for a newly generated packet is granted only at the start of the next frame (see Fig. \ref{fig:irsa_timeline}). 
For the traffic generation model under study, this implies that a node may become active more than once over an \slots-slot period, e.g. being triggered multiple times by changes in the physical quantity it monitors. Recalling that the goal of the network is to provide the destination with the most up-to-date state information, a device in such situation accesses the channel at the beginning of the subsequent frame to transmit the last generated update only, discarding all the others.\footnote{In other words, each node can be thought of as a one-packet sized buffer, with preemption by fresher updated allowed only in waiting.} As for \ac{SA}, neither feedback from the sink nor retransmission strategy is considered.

Following this model, the number \actNodesFrame\ of devices that initiate a transmission at the start of a frame is binomially distributed, i.e. $\actNodesFrame \sim \text{Bin}(\nodes,1-(1-\pAct)^\slots)$, where the second parameter captures the probability for a device to become active at least once over the previous frame.
Accordingly, the channel load \load, defined as the average number of users which transmit per slot, evaluates to
\begin{equation}
    \load := \frac{\expOp[\actNodesFrame]}{\slots} = \frac{\nodes \left( 1 - (1-\pAct)^\slots \right)}{\slots}.
    \label{eq:load_irsa}
\end{equation}

As per the IRSA protocol, each of the $\actNodesFrame$ nodes transmits \replicas\ copies of its status update, uniformly placed at random over the \slots\ available slots. Each packet contains a pointer\footnote{We also mention another approach proposed in \cite{vem2017user}. The idea is that the message itself is the seed of a random generator, used by the node to place its replicas. As soon as a packet is decoded, the receiver can then use the seed to retrieve the positions of the twins, eliminating the need for additional overhead to store the pointers.} to the position of its replicas, and the number of copies is independently drawn at the start of the frame from the distribution 
\begin{equation}
    \Lambda(x) = \sum_{\ell=1}^L \Lambda_{\ell} \, x^\ell
\end{equation}
where $\Lambda_\ell$ indicates the probability of sending $\ell$ copies. 
At the receiver side the whole frame is stored, and the decoding process is initiated by looking for collision-free slots. For any such time unit, the corresponding update is decoded, and the interference contribution of all its replicas is removed from the frame, possibly leading to more singleton \--- and decodeable \--- slots. The procedure is iterated until all users have been retrieved or no more slots with a single packet can be found. 

An illustration of \ac{IRSA} operation is reported in Fig.~\ref{fig:irsa_timeline}, considering the simple case of $\nodes= 4$ users and $\slots=6$ slots per frame. During frame $k$ (Fig.~\ref{fig:irsa_timeline_tx}), each node randomly becomes active at any slot with probability \pAct. In the example, for instance, user $1$ generates updates at slots $2$ and $4$, while user $3$ never activates. Accordingly, during the successive frame (Fig.~\ref{fig:irsa_timeline_rx}), the latter will not transmit, whereas the former will access the channel to send a packet. To do so, the user draws the number of replicas to transmit (e.g., $2$ for user $1$), and places them uniformly over the available slots. Each packet copy contains the same time-stamp, marking the instant at which the freshest update was generated by the terminal over the previous frame (e.g., $(k-1)\slots + 3$ in the case of user $1$, as the node was last awaken at the end of the third slot). At the receiver, once the whole $(k+1)$-th frame is stored, decoding starts by retrieving the third copy of user $2$'s packet, which was received in a singleton slot. Removing the other two replicas renders the second copy of user $4$ decodeable. Finally, once the replicas of this information piece are cleared as well, user $1$ is also perceived as interference free and can be retrieved. 

In terms of medium access efficiency, the throughput of \ac{IRSA} can be generally expressed in the form \mbox{$\tru = (1-\ploss) \cdot \load$}, where $(1-\ploss)$ indicates the probability for a transmitted information unit to be decoded at the sink. In contrast to \ac{SA}, no close-form expression for the packet loss rate \ploss\ is known to date, yet some approximation for finite length frames can be derived, as will be discussed in more details in Sec.~\ref{sec:irsa_plr}.

As to \ac{AoI}, we assume decoding to start once the whole frame is buffered, and processing time for \ac{SIC} operations to be negligible, so that all updates are elaborated by the destination at the end of the \slots-slot period. Upon successful decoding of a user, then, its current age is reset to the sum of a frame duration \--- needed to transmit and retrieve the message \--- and of the time elapsed from the update generation to the start of the frame it was transmitted on. With reference to the situation reported in Fig.~\ref{fig:irsa_timeline}, for instance, at the end of the $(k+1)$-th frame we have $\age^{(1)}((k+1)\slots) = \slots + 3$.

\section{Analysis}
\label{sec:analysis}

\begin{figure*}[t!]
\normalsize
\setcounter{equation}{5}
\begin{align}
\begin{split}
\Agei &= \lim_{\ell\rightarrow\infty}  \frac{\ell}{\sum_{j=1}^{\ell} \slots Z_j} \left( \frac{\mathcal A_1}{\ell} +
    \frac{1}{\ell} \sum_{j=2}^{\ell} \left( \mathcal A_j' + \mathcal A_j'' \right)
    \right)\\
    &\stackrel{(a)}{=} \lim_{\ell\rightarrow\infty} \frac{\ell}{\sum_{j=1}^\ell Z_j} \left( \frac{\slots}{\ell} \sum_{j=2}^\ell Z_j +  \frac{\slots}{2 \ell} \sum_{j=2}^\ell Z_j^2 + \frac{1}{\ell}\sum_{j=2}^\ell Z_j X_{j-1} \right)
    \stackrel{(b)}{=} \frac{1}{\expOp[Z_j]} \left( \slots \expOp[Z_j] +  \frac{\slots}{2} \expOp[Z_j^2] + \expOp[Z_j X_{j-1}] \right)
\end{split}
\label{eq:aoi_steps}
\end{align}
\hrulefill
\end{figure*}
\setcounter{equation}{3}
The key contribution of our analysis is summarised by the following result:
\begin{prop} \label{prop:aoi_irsa}
For an \ac{IRSA} scheme, under the traffic model described in Sec.~\ref{sec:sysModel}, the average network age of information expressed in slots is given by
\begin{equation}
    \AoIIRSA = \frac{\slots}{2} + \frac{\nodes}{\tru} + \left( \frac{1}{\pAct} - \frac{\slots(1-\pAct)^{\slots}}{1 - (1-\pAct)^{\slots}} \right).
    \label{eq:aoi_irsa}
\end{equation}
\end{prop}
\begin{IEEEproof}
Consider the evolution of the current \ac{AoI} $\agei(t)$ at a generic node $i$ reported in Fig. \ref{fig:timeline_AoI}. 
Without loss of generality, we assume $\agei(0) = 0$, and track the system behaviour over a period corresponding to an integer number of frames, coming to an end with the $\ell$-th reception of a packet from terminal $i$. 
Let us now focus on the $j$-th \emph{successfully delivered} update, and define as $X_{j}$  the number of slots elapsed between its generation and the start of the next frame (i.e. when the packet transmission is initiated).\footnote{We remark that packets transmitted but not correctly received are not relevant for the \ac{AoI} computation, and their generation times are not tracked.} Moreover, denote as $Z_j$ the number of frames elapsed between retrieval of the $(j-1)$-th and the $j$-th updates at the sink. As a first remark, we observe that the no-retransmission policy under study, as well as the independent behaviour of nodes within the network and across frames, render both the random processes $\{X_j\}$ and $\{Z_j\}$ i.i.d. and independent between each other.
\begin{figure}
    \centering
    \includegraphics[width=.98\columnwidth]{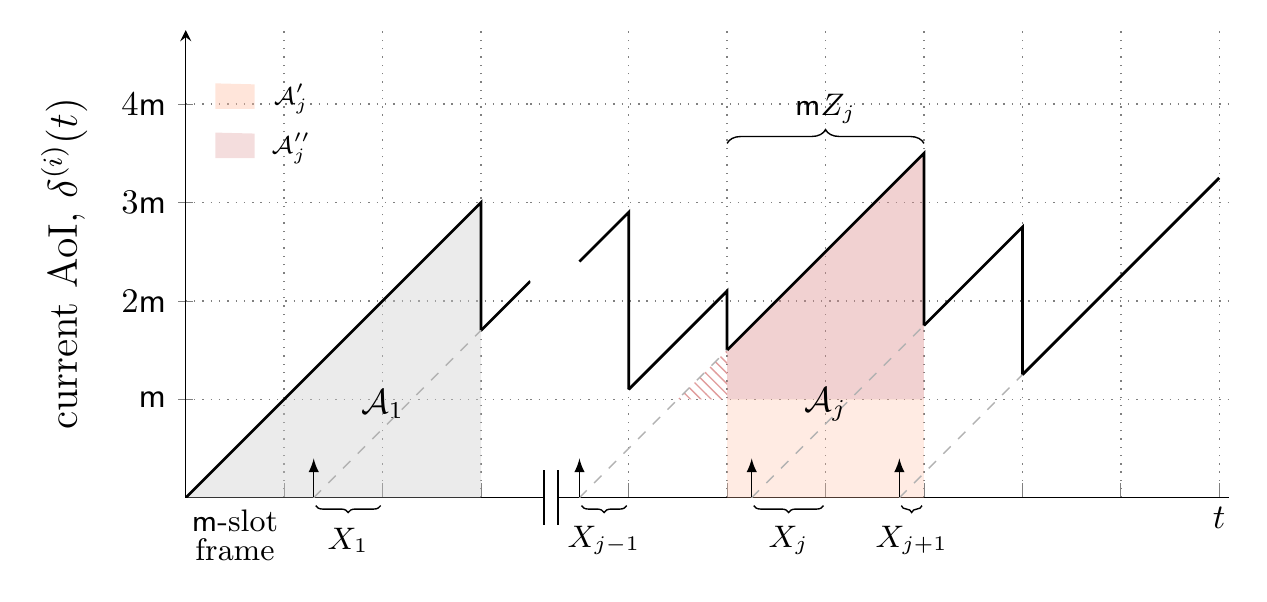}
    \vspace{-.8em}
    \caption{Example timeline for the AoI evolution at node $i$.}
    \label{fig:timeline_AoI}
    \vspace{-1.1em}
\end{figure}
Furthermore, following this notation, we have
\begin{equation}
\tau = \sum_{j=1}^\ell \slots Z_j.
\end{equation} 
If we now indicate as $\mathcal A_j$ the area under the curve $\agei(t)$ within the $Z_j$ frames, the integral in \eqref{eq:def_av_aoi_node} can be expressed as a sum of components, leading a convenient formulation for the terminal's average \ac{AoI}:
\begin{equation}
    \Agei = \lim_{\ell \rightarrow \infty} \frac{1}{\sum_{j=1}^\ell \slots Z_j} \cdot \sum_{j=1}^{\ell} \mathcal A_j.
    \label{eq:av_aoi_sum}
\end{equation}

In turn, the trapezoidal area $\mathcal A_j$ can be written for any $j>1$ as the sum of the two contributions $\mathcal A_j'$ and $\mathcal A_j''$ (see  Fig~\ref{fig:timeline_AoI}). The former, a rectangle with sides of length $\slots Z_j$ and $\slots$, accounts for the fact that updates can only be received at the end of a frame. The latter, instead, can be computed as the difference between the areas of two isosceles right triangles of equal sides $(\slots Z_j + X_{j-1})$ and $X_{j-1}$, respectively. On the other hand, for $j=1$, we simply have $\mathcal A_1=(\slots Z_1)^2 /2$, where $Z_1$ indicates the number of frames elapsed before reception of the first update from the node. Plugging these results into \eqref{eq:av_aoi_sum} and performing simple algebraic manipulations, we obtain the expression in \eqref{eq:aoi_steps}, reported at the top of the page.
\setcounter{equation}{6}
Here, (a) stems from the observation that the term $\mathcal A_1/\ell$ becomes negligible as $\ell$ grows to infinity, while (b) follows by the law of large numbers for the involved i.i.d. processes. 
Finally, invoking the independence of $Z_j$ and $X_{j-1}$, we obtain
\begin{equation}
    \Agei = \slots +  \frac{\slots}{2} \cdot \frac{\expOp[Z_j^2]}{\expOp[Z_j]} + \expOp[X_{j-1}].
    \label{eq:aoi_expectations}
\end{equation}
\vspace{-.3em}
The average \ac{AoI} of a node can thus determined by simply computing the first and second order moments of two r.v.s.
To this aim, consider first $Z_j$. For the system model under study, every frame sees two possible outcomes for a node: either the receiver refreshes the \ac{AoI} having decoded a terminal's update, or no message from the node is retrieved. Moreover, given that no retransmissions are performed in case of delivery failure and that all devices operate independently, the outcomes of successive frames are i.i.d. The number of frames between two successful updates for a terminal, $Z_j$, follows thus a geometric distribution with parameter $\nu := (1-\ploss) \cdot \, \left[1-(1-\pAct)^\slots \right]$, where $(1-\ploss)$ captures the probability of decoding a packet achieved by the employed \ac{IRSA} access scheme, and the second factor accounts for the probability of the node having an update to transmit at the start of the frame. Accordingly, the first and second order moments of $Z_j$ evaluate to
\begin{align}
    \expOp\left[ Z_j \right] = \frac{1}{\nu} \hspace{2em}
    \expOp\left[ Z_j^2 \right] = \frac{2 - \nu}{\nu^2}.
    \label{eq:moments_Zj}
\end{align}
Moreover, recalling the definition of channel load in \eqref{eq:load_irsa}, $\nu$ can be conveniently expressed as $\nu = \tru \cdot \slots/\nodes$.

Let us now instead focus on $X_j$. The r.v. has alphabet  $\{1,\dots,\slots\}$ and, by definition, takes value $k$ when the node became active for the \emph{last} time on slot $\slots-k+1$ during the frame preceding a successful update. In other words, 
\begin{equation}
    \pr \left\{ X_j = k \right\} = \frac{\pAct (1-\pAct)^{k-1}}{1 - (1-\pAct)^\slots}.
    \label{eq:pX}
\end{equation}
Within \eqref{eq:pX}, the numerator accounts for the probability of the node becoming active at slot $\slots-k+1$ and then remaining inactive until the end of the frame ($k-1$ slots), i.e. of not generating fresher updates that would supersede the one being considered. The denominator, instead, provides the normalisation condition capturing the fact that at least one update has to be generated over the $\slots$-slot period for the node to be successfully received during the subsequent frame. Leaning on this, the expected value for $X_j$ follows after simple manipulations:
\begin{equation}
    \expOp \left[ X_j \right] = \sum_{k=1}^{\slots} k  \cdot \pr \left\{ X_j = k \right\} = \frac{1}{\pAct} - \frac{\slots (1-\pAct)^\slots}{1-(1-\pAct)^\slots}.
    \label{eq:moment_Xj}
\end{equation}

Plugging \eqref{eq:moments_Zj} and \eqref{eq:moment_Xj} into \eqref{eq:aoi_expectations}, the average \ac{AoI} for the generic terminal $i$ evaluates to the right-handside of \eqref{eq:aoi_irsa}. Recalling that all nodes in the system operate independently, we have $\AoI = \Agei$, which proves the proposition.
\end{IEEEproof}

Leaning on the derivation of Prop. \ref{prop:aoi_irsa}, it is also useful to state the following result, in agreement with \cite{Yates17:AoI_SA,Modiano18_AoI}:
\begin{lemma} \label{lemma_aoiSA}
For a \ac{SA} scheme, under the traffic model described in Sec.~\ref{sec:sysModel}, the average network age of information expressed in slots is given by
\begin{equation}
    \AoISA = \frac{1}{2} + \frac{\nodes}{\tru}.
    \label{eq:aoi_sa}
\end{equation}
\end{lemma}
\begin{IEEEproof}
The statement follows by observing how the analysis carried out in the proof of Prop.~\ref{prop:aoi_irsa} captures the behaviour of \ac{SA} when $\slots = 1$.
\end{IEEEproof}

The exact closed-form expression derived in \eqref{eq:aoi_irsa} offers a compact characterisation of the average network \ac{AoI} when \ac{IRSA} is employed. Notably, \AoIIRSA\ is given by the sum of three components, which provide insight on the system behaviour. The first, $\slots/2$, accounts for the linear evolution of the information age during the one $\slots$-slot frame needed for successfully transmitting and decoding an update. Moreover, recalling \eqref{eq:moments_Zj} we readily get $\nodes/\tru = \slots \expOp[Z_j]$, so that the second addend in \eqref{eq:aoi_irsa} accounts for the age increment due to the average number of slots elapsed between two successful updates. Finally, the third component of \AoIIRSA\ quantifies the additional latency induced by frame-based operations, capturing the average number of slots between the generation of an update and the start of its transmission, i.e. $\expOp[X_j]$. 

This interpretation prompts a first interesting comparison with the performance of \ac{SA}. Indeed, an inspection of \eqref{eq:aoi_sa} highlights how the simpler protocol benefits from its slot-by-slot operation, enjoying a shorter average transmission time \--- $1/2$ in place of $\slots/2$ \---, and not being beset by additional waiting time between node activation and transmission. Conversely, the \ac{AoI} dependency on throughput takes the same form in both cases, favouring the better performance offered by IRSA. From this standpoint, thus, a key tradeoff between frame duration and transmission efficiency arises, which will be thoroughly discussed in Sec.~\ref{sec:results}.

\subsection{A throughput approximation for IRSA} \label{sec:irsa_plr}
Computation of the exact expression for \AoIIRSA\ reported in \eqref{eq:aoi_irsa} requires knowledge of the aggregate throughput \mbox{$\tru = (1-\ploss) \load$}, for which, to date, a closed form expression as a function of the channel load remains elusive. On the other hand, recent research efforts have identified tight analytical approximations of the packet loss rate for finite length frames under the destructive collision model discussed in Sec.~\ref{sec:sysModel}. Specifically, the behaviour of IRSA in the \emph{error-floor} region, i.e. at low channel load, was characterised in \cite{Sandgren17_TCOM,Ivanov17_TCOM}, whereas accurate expressions for \ploss\ in the moderate to high channel load regime (\emph{waterfall region}) were derived in \cite{Graell18_Waterfall},  adapting the finite length scaling analysis of low-density parity check codes originally proposed in \cite{Urbanke09_TIT}.

\begin{figure}
    \centering
    \includegraphics[width=.9\columnwidth]{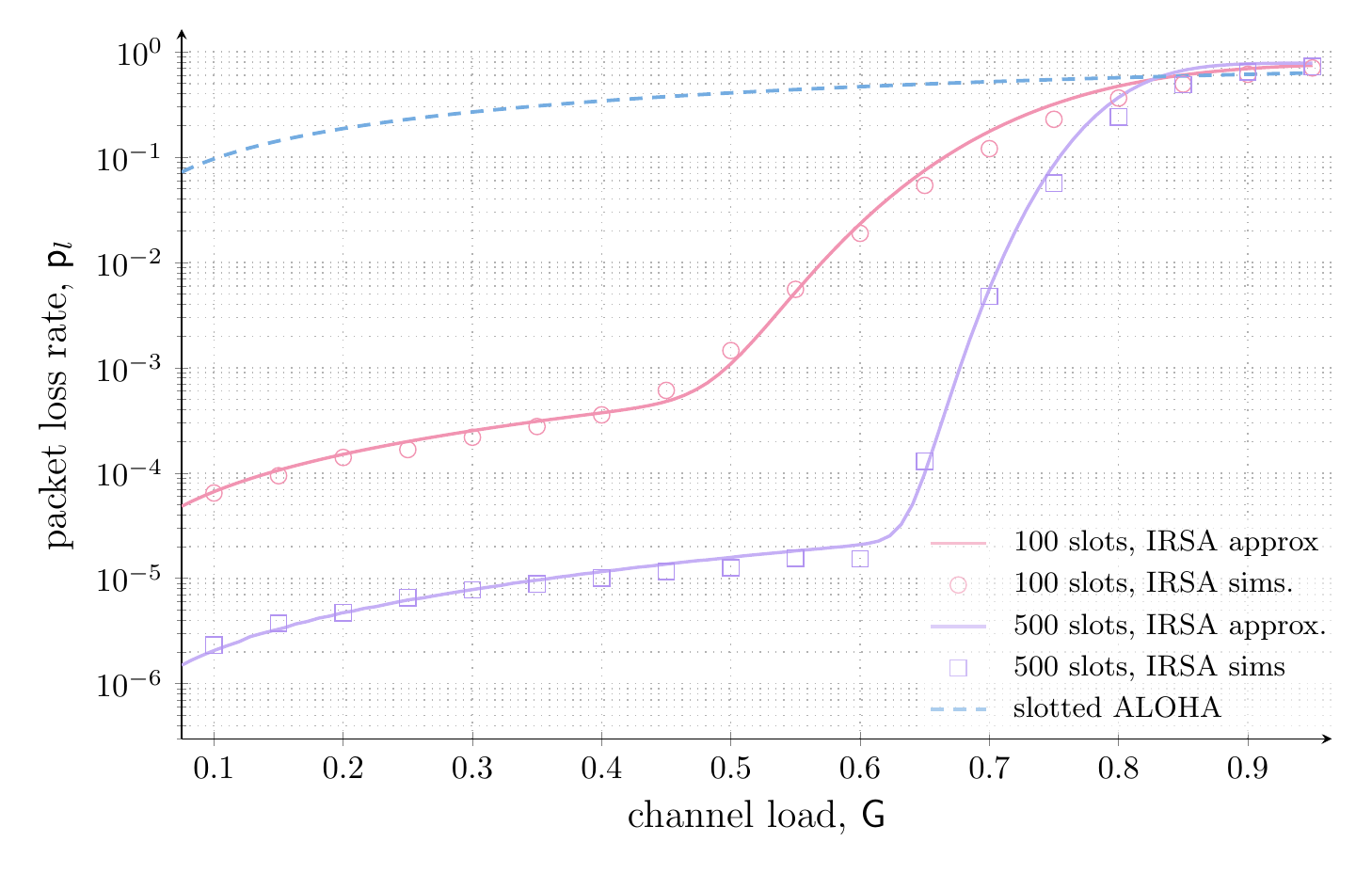}
    \vspace{-.8em}
    \caption{Packet loss rate $\ploss$ vs channel load \load. Solid lines for IRSA ($\Lambda = x^3$) report analytical expressions obtained via the approximation \eqref{eq:plr_approx}, whereas markers indicate simulation results.}
    \label{fig:plr_irsa}
    \vspace{-.8em}
\end{figure}

Let us denote these analytical approximations, not explicitly reported here due to space constraints, as $\plossef(\load)$ and $\plosswf(\load)$, respectively. Despite being defined for any value of \load, each quantity captures effects that crucially impact performance in one of the two load regions, having little effect on the other one. As a result, $\plosswf \ll \plossef$ in the error-floor region and, similarly, $\plossef \ll \plosswf$ in the waterfall region. Leaning on this observation, we consider in this work a simple approximation of the packet loss rate in the form
\begin{equation}
    \ploss(\load) \simeq \plossef(\load) + \plosswf(\load)
    \label{eq:plr_approx}
\end{equation}
which allows to easily derive analytical estimates of the \ac{AoI} for IRSA for any configuration of interest (i.e., for any $\nodes$, $\slots$ and \pAct). The tightness of the expression is shown in Fig.~\ref{fig:plr_irsa} for the reference case $\Lambda(x) = x^3$, reporting the packet loss rate against the channel load for two distinct frame lengths obtained via  \eqref{eq:plr_approx} (solid lines) and by mean of simulations (markers). The plot also highlights how operating \ac{IRSA} over longer frames triggers better performance, lowering the error floor and shifting the threshold for entering the waterfall region to higher channel loads \cite{Sandgren17_TCOM,Graell18_Waterfall}.

\section{Results and Discussion}
\label{sec:results}

\begin{figure}
    \centering
    \includegraphics[width=.9\columnwidth]{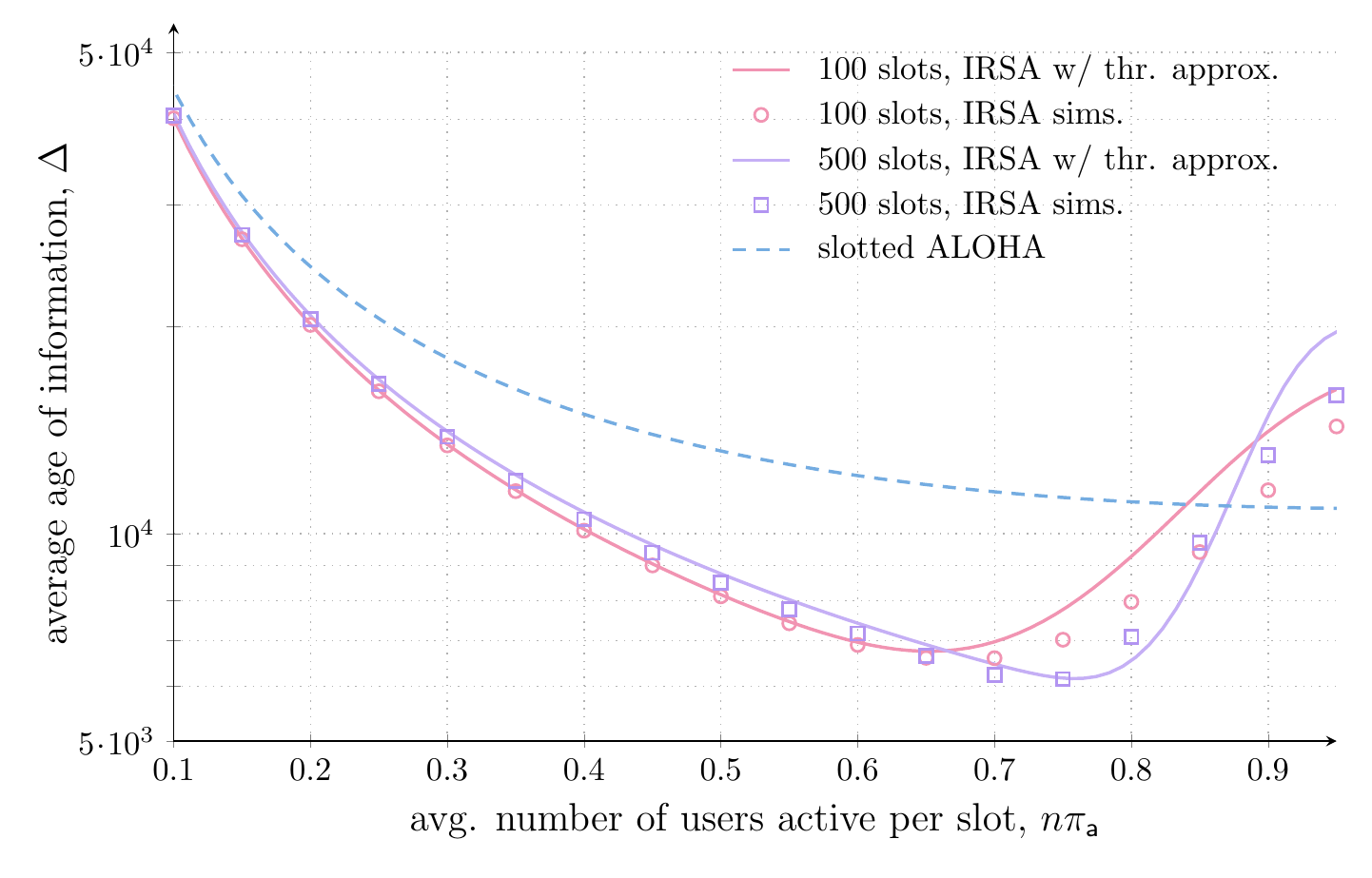}
    \vspace{-.8em}
    \caption{Average network \ac{AoI} \AoI\ vs. average number of active users per slot $\nodes\pAct$. Solid lines report analytical trends obtained evaluating \eqref{eq:aoi_irsa} via the throughput approximation of Sec.~\ref{sec:irsa_plr}, while markers the outcome of simulations. Results obtained for $\nodes=4000$ users.}
    \label{fig:ageVsLoad}
\end{figure}

To draw insights on the behaviour of different random access policies, we start by studying \AoI\ as a function of the average number of users per slot that generate an update ($n\pAct$). Unless otherwise stated, we focus on a system with $\nodes = 4000$ terminals, and assume a distribution $\Lambda(x) = x^3$ for \ac{IRSA} operations (i.e., a node accessing the channel sends three copies of its packet over the frame). The results obtained in this configuration are reported in Fig.~\ref{fig:ageVsLoad}, together with the performance of \ac{SA} (dashed line). 

All access policies exhibit a common trend as nodes' activity increases, characterised by a point of minimum for the average \ac{AoI} experienced at moderate-to-high channel load. Indeed, while a lightly loaded medium grants high chance of success for a transmitted update, the behaviour for low values of $\nodes \pAct$ is driven by the scarce activity of terminals, resulting in long inter-update generation times (low \pAct) that penalise \AoI. Conversely, too frequent reporting from nodes ($\nodes \pAct \sim 1$) lead to channel congestion, hindering decoding of information at the receiver due to collisions. For \ac{SA}, an inspection of \eqref{eq:aoi_sa} readily reveals how \--- for a given terminal population \nodes\ \--- the minimum \ac{AoI} $\AoISA^*$ is achieved when the aggregate throughput \tru\ is maximised. This happens for $\pAct = 1/\nodes$, leading to
\begin{equation}
    \AoISA^* = \frac{1}{2} + \nodes \left(1-\frac{1}{\nodes}\right)^{1-\nodes} \simeq
    \frac{1}{2} + \nodes e
    \label{eq:optAoI_SA}
\end{equation}
where the approximation quickly becomes very tight for large enough, and practical, values of \nodes. As to \ac{IRSA}, instead, changes in the activation probability \pAct\ affect not only the throughput, but also the additional term $\expOp[X_j]$ accounting for the latency between update generation and transmission, prompting a more convoluted dependency of $\AoIIRSA^*$ on \nodes\pAct.

More interestingly, Fig.~\ref{fig:ageVsLoad} highlights how \ac{IRSA} consistently outperforms \ac{SA} in terms of average \ac{AoI} for most values of \nodes\pAct, and triggers stark improvements exactly in the moderate load conditions under which many practical systems are operated. Such a result is non-trivial, as it clarifies how the benefits in terms of throughput efficiency offered by repetitions and \ac{SIC} do outweigh the additional latency cost induced by framed channel access. A first relevant design hint is thus offered, suggesting the use of modern random access solutions as a means to reduce \ac{AoI} in grant-free based \ac{mMTC} applications.\footnote{The analytical approximation of Sec.~\ref{sec:irsa_plr} tightly predicts performance in the region of most interest, i.e. up to the point of minimum for \AoIIRSA. Agreement with simulations slightly deteriorates for larger values of $\nodes\pAct$ \---  despite the exact formulation in \eqref{eq:aoi_irsa} \--- due to the less accurate approximation of \tru\ in the waterfall region \cite{Graell18_Waterfall}.}

A second key message emerges when comparing the performance of \ac{IRSA} for different frame sizes. When few nodes per slot become active, the use of shorter frames (e.g. $\slots = 100$) is beneficial, as the system operates in the \emph{error floor} region, where very low packet loss rates trigger small throughput improvements when increasing \slots. In this case, the average latency terms due to time between update generation and channel access $(\expOp[X_j])$, and transmission ($\slots/2$), drive \AoIIRSA. Conversely, for larger values of \nodes\pAct, the throughput improvements brought by operating \ac{IRSA} over longer frames become dominant. From this standpoint, not only does performance improve for moderate channel loads, but also the minimum achievable average \ac{AoI} $\AoIIRSA^*$ experiences a reduction.

\begin{figure}
    \centering
    \includegraphics[width=.9\columnwidth]{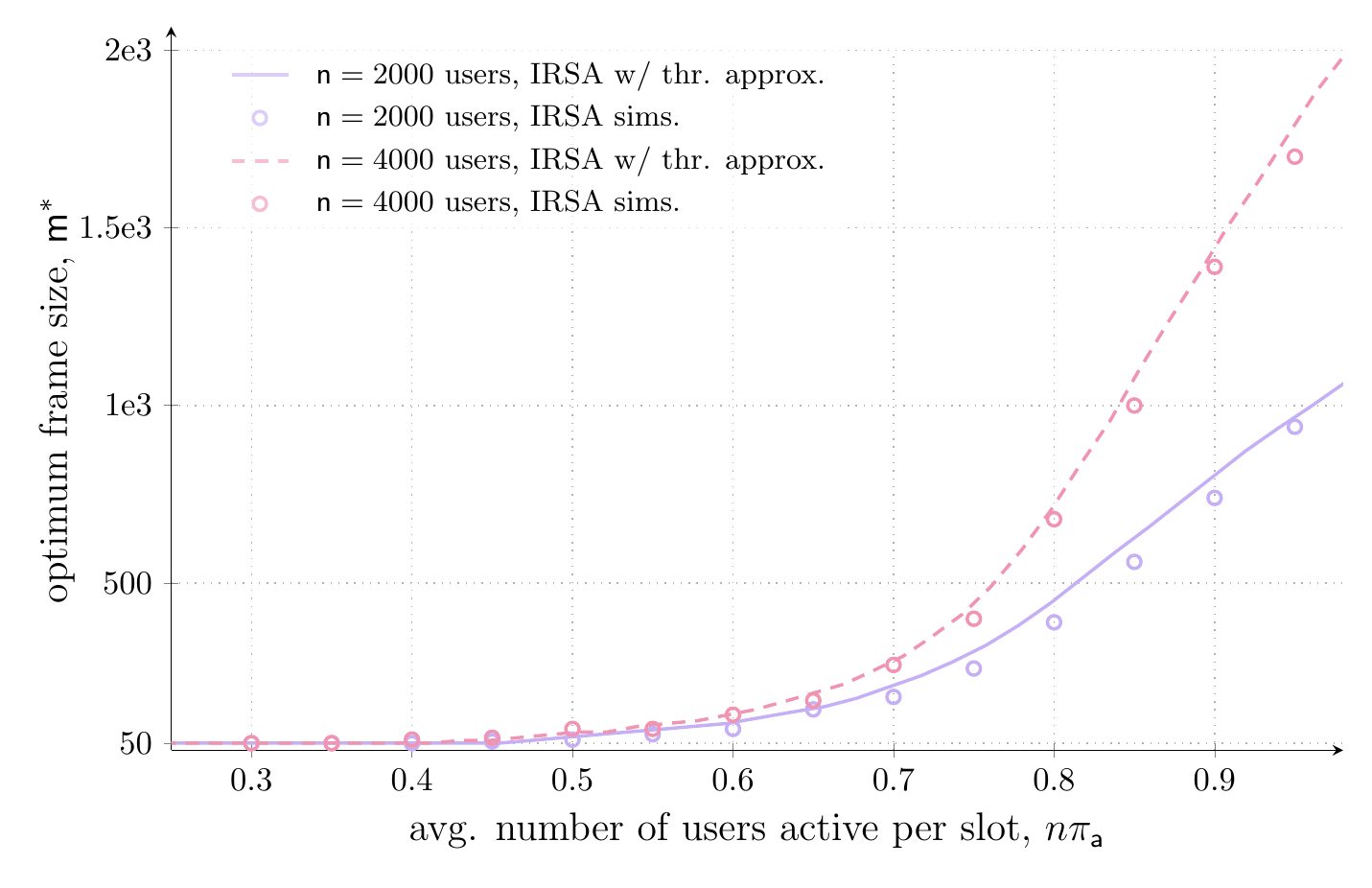}
    \vspace{-.8em}
    \caption{Optimum IRSA frame size $\slots^*$ vs average number of active users per slot for $\nodes=2000$ and $\nodes=4000$.}
    \label{fig:optFrameSize}
    \vspace{-1em}
\end{figure}

\begin{figure}
    \centering
    \includegraphics[width=.9\columnwidth]{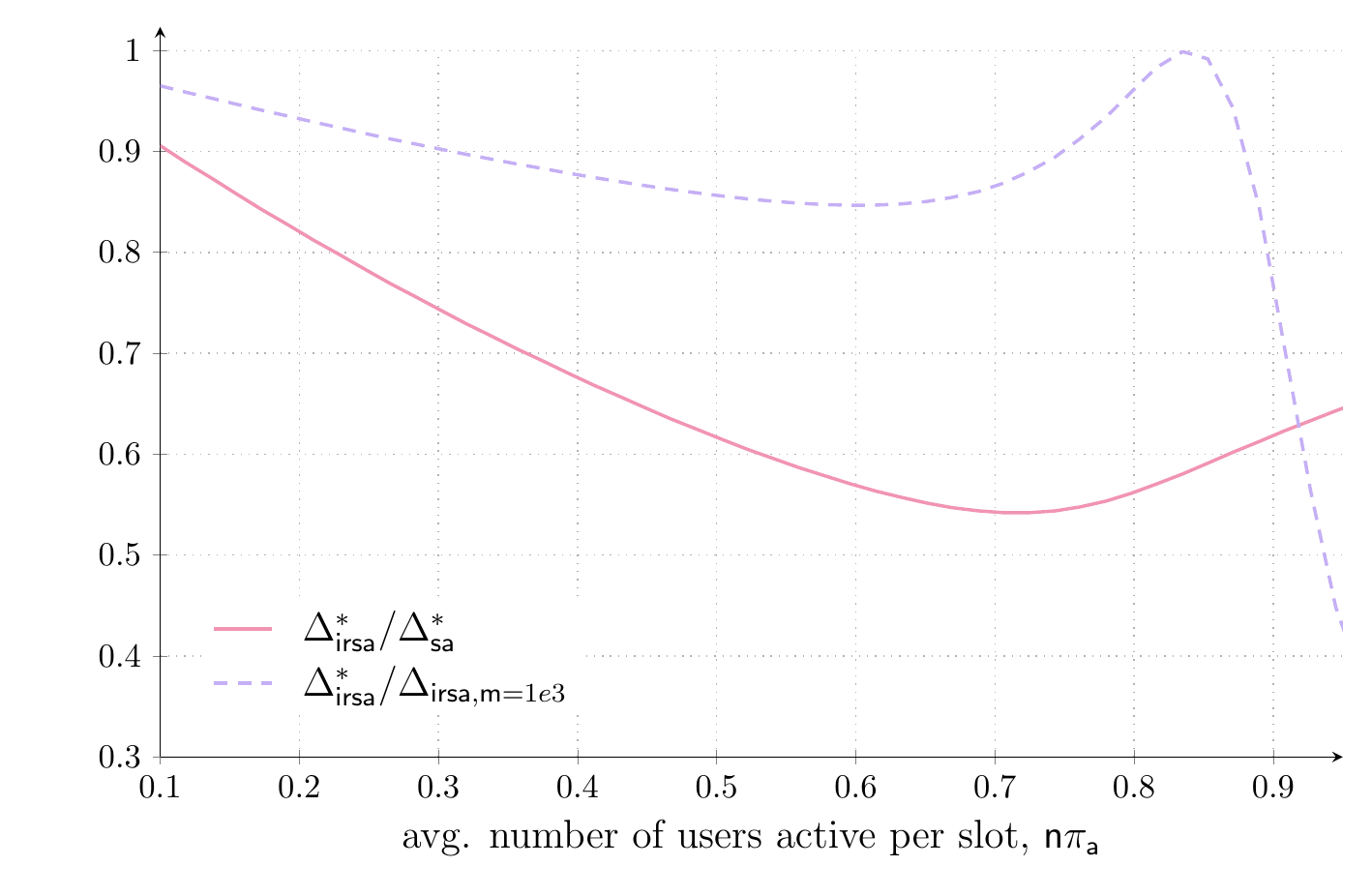}
    \vspace{-.8em}
    \caption{Ratio of \ac{AoI} for IRSA when operated at optimum frame size $\slots^*$ and at fixed $\slots=1000$ frame size (dashed line), and ratio of \ac{AoI} for IRSA operated at $\slots^*$ frame size and SA (solid line). In all cases, $\nodes=4000$.}
    \label{fig:ratios}
    \vspace{-1em}
\end{figure}

This observation pinpoints a fundamental tradeoff between frame length and \ac{AoI}, and triggers the natural question on how to properly select \slots\ given an average terminal activity pattern in order to optimise information freshness. The issue is explored in Fig.~\ref{fig:optFrameSize}, which shows the frame size $\slots^*$ minimising \AoIIRSA\ for any value of \nodes\pAct. As discussed, short frames are to be preferred in lightly loaded channel conditions, whereas $\slots^*$ shall be carefully tuned to provide optimal performance when $\nodes\pAct$ is increased. In all setups, the characterisation of \AoIIRSA\ derived in \eqref{eq:aoi_irsa} offers a useful and handy system design tool. 
To better gauge the impact of frame size, we report in Fig.~\ref{fig:ratios} two further quantities of interest. The dashed line describes the ratio of the minimum achievable \ac{AoI} $\AoIIRSA^*$ (obtained when operating \ac{IRSA} with frame size $\slots^*$) to the average \ac{AoI} \AoIIRSA\ experienced when a fixed frame size $\slots=1000$ is employed, irrespective of \nodes\pAct. A remarkable impact emerges, as a proper choice of \slots\ can offer improvements of up to $15\%$ at low-to-intermediate channel conditions, and becomes paramount for higher values of \nodes\pAct. Similarly, the solid line shows the ratio of $\AoIIRSA^*$ to $\AoISA^*$, i.e. to the minimum \ac{AoI} obtained by \ac{SA} and discussed in \eqref{eq:optAoI_SA}. Once again, the benefits of using modern random access schemes are apparent, with an \ac{AoI} almost halved for channel loads of practical interest.

\bibliographystyle{IEEEtran}
\bibliography{IEEEabrv,aloha}

\begin{thebibliography}{10}
\providecommand{\url}[1]{#1}
\csname url@samestyle\endcsname
\providecommand{\newblock}{\relax}
\providecommand{\bibinfo}[2]{#2}
\providecommand{\BIBentrySTDinterwordspacing}{\spaceskip=0pt\relax}
\providecommand{\BIBentryALTinterwordstretchfactor}{4}
\providecommand{\BIBentryALTinterwordspacing}{\spaceskip=\fontdimen2\font plus
\BIBentryALTinterwordstretchfactor\fontdimen3\font minus
  \fontdimen4\font\relax}
\providecommand{\BIBforeignlanguage}[2]{{%
\expandafter\ifx\csname l@#1\endcsname\relax
\typeout{** WARNING: IEEEtran.bst: No hyphenation pattern has been}%
\typeout{** loaded for the language `#1'. Using the pattern for}%
\typeout{** the default language instead.}%
\else
\language=\csname l@#1\endcsname
\fi
#2}}
\providecommand{\BIBdecl}{\relax}
\BIBdecl

\bibitem{Kaul11_SECON}
S.~{Kaul}, M.~{Gruteser}, V.~{Rai}, and J.~{Kenney}, ``Minimizing age of
  information in vehicular networks,'' in \emph{Proc. IEEE SECON}, June 2011.

\bibitem{Kaul11_Globecom}
S.~{Kaul}, R.~{Yates}, and M.~{Gruteser}, ``On piggybacking in vehicular
  networks,'' in \emph{Proc. IEEE GLOBECOM}, Dec 2011.

\bibitem{Yates19_TIT}
R.~D. {Yates} and S.~K. {Kaul}, ``The age of information: Real-time status
  updating by multiple sources,'' \emph{IEEE Transactions on Information
  Theory}, vol.~65, no.~3, pp. 1807--1827, March 2019.

\bibitem{Durisi19_JSAC}
R.~{Devassy}, G.~{Durisi}, G.~C. {Ferrante}, O.~{Simeone}, and E.~{Uysal},
  ``Reliable transmission of short packets through queues and noisy channels
  under latency and peak-age violation guarantees,'' \emph{{IEEE} J. Sel. Areas
  Commun.}, vol.~37, no.~4, pp. 721--734, April 2019.

\bibitem{Modiano19_book}
Y.~{Sun}, I.~{Kadota}, R.~{Talak}, E.~{Modiano}, and R.~{Srikant}.\hskip 1em
  plus 0.5em minus 0.4em\relax Morgan \& Claypool, 2019.

\bibitem{Yates17:AoI_SA}
R.~Yates and S.~Kaul, ``Status updates over unreliable multiaccess channels,''
  in \emph{Proc. IEEE ISIT}, June 2017.

\bibitem{Modiano18_AoI}
R.~Talak, S.~Karaman, and E.~Modiano, ``Distributed scheduling algorithms for
  optimizing information freshness in wireless networks,'' in \emph{Proc. IEEE
  SPAWC}, June 2018.

\bibitem{Ephremides19_Infocom}
A.~{Maatouk}, M.~{Assaad}, and A.~{Ephremides}, ``Minimizing the age of
  information: Noma or oma?'' in \emph{Proc. IEEE INFOCOM Workshops}, April
  2019.

\bibitem{Ephremides20_CSMA}
------, ``On the age of information in a csma environment,'' \emph{IEEE/ACM
  Transactions on Networking}, 2020.

\bibitem{Yates20_arXiv}
\BIBentryALTinterwordspacing
R.~Yates, , and S.~Kaul, ``{Age of Information in Uncoordinated Unslotted
  Updating},'' 2020. [Online]. Available: \url{http://arxiv.org/abs/2002.02026}
\BIBentrySTDinterwordspacing

\bibitem{Shirin19_arXiv}
\BIBentryALTinterwordspacing
X.~Chen, K.~Gatsis, H.~Hassani, and S.~Bidokhti, ``{Age of Information in
  Random Access Channels},'' 2020. [Online]. Available:
  \url{http://arxiv.org/abs/1912.01473}
\BIBentrySTDinterwordspacing

\bibitem{Soung20_arXiv}
\BIBentryALTinterwordspacing
H.~Chen, Y.~Gu, and S.-C. Liew, ``{Age-of-Information Dependent Random Access
  for Massive IoT Networks},'' 2020. [Online]. Available:
  \url{http://arxiv.org/abs/2001.24780}
\BIBentrySTDinterwordspacing

\bibitem{Berioli2016}
M.~Berioli, G.~Cocco, G.~Liva, and A.~Munari, \emph{{Modern random access
  protocols}}.\hskip 1em plus 0.5em minus 0.4em\relax NOW Publisher, 2016.

\bibitem{Paolini15:TIT_CSA}
E.~Paolini, G.~Liva, and M.~Chiani, ``{Coded Slotted ALOHA: A Graph-Based
  Method for Uncoordinated Multiple Access},'' \emph{{IEEE} Trans. Inf.
  Theory}, vol.~61, no.~12, pp. 6815--6832, 2015.

\bibitem{Liva11:IRSA}
G.~Liva, ``{Graph-Based Analysis and Optimization of Contention Resolution
  Diversity Slotted {ALOHA}},'' \emph{{IEEE} Trans. Commun.}, vol.~59, no.~2,
  pp. 477--487, 2011.

\bibitem{Abramson77:PacketBroadcasting}
N.~Abramson, ``{The Throughput of Packet Broadcasting Channels},'' \emph{{IEEE}
  Trans. Commun.}, vol. COM-25, no.~1, pp. 117--128, 1977.

\bibitem{vem2017user}
A.~Vem, K.~R. Narayanan, J.~Cheng, and J.-F. Chamberland, ``{A user-independent
  serial interference cancellation based coding scheme for the unsourced random
  access Gaussian channel},'' in \emph{Information Theory Workshop (ITW), 2017
  IEEE}.\hskip 1em plus 0.5em minus 0.4em\relax IEEE, 2017, pp. 121--125.

\bibitem{Sandgren17_TCOM}
E.~{Sandgren}, A.~{Graell i Amat}, and F.~{Brännström}, ``On frame
  asynchronous coded slotted aloha: Asymptotic, finite length, and delay
  analysis,'' \emph{IEEE Trans. Commun.}, vol.~65, no.~2, pp. 691--704, 2017.

\bibitem{Ivanov17_TCOM}
M.~{Ivanov}, F.~{Brännström}, A.~{Graell i Amat}, and P.~{Popovski},
  ``Broadcast coded slotted aloha: A finite frame length analysis,'' \emph{IEEE
  Trans. Commun.}, vol.~65, no.~2, pp. 651--662, 2017.

\bibitem{Graell18_Waterfall}
A.~{Graell i Amat} and G.~{Liva}, ``Finite-length analysis of irregular
  repetition slotted aloha in the waterfall region,'' \emph{IEEE Commun.
  Letters}, vol.~22, no.~5, pp. 886--889, 2018.

\bibitem{Urbanke09_TIT}
A.~{Amraoui}, A.~{Montanari}, T.~{Richardson}, and R.~{Urbanke},
  ``Finite-length scaling for iteratively decoded ldpc ensembles,'' \emph{IEEE
  Trans. Inf. Theory}, vol.~55, no.~2, pp. 473--498, 2009.

\end{thebibliography}

\end{document}